\documentclass{PoS}
\usepackage{xcolor}
\usepackage{wrapfig}
\usepackage{amsmath}
\usepackage{physics}
\renewenvironment{thebibliography}[1]{
  \begin{oldthebibliography}{#1}
    \setlength{\itemsep}{0.4em}
  \setlength{\parskip}{0em}
}
{
 \end{oldthebibliography}
}

\title{Eleven Year Search for Supernovae with the IceCube Neutrino Observatory}

\DeclareMathOperator{\Var}{var}

\ShortTitle{Supernovae in IceCube}

\author{
The IceCube Collaboration\footnote{For collaboration list, see PoS(ICRC2019) 1177.}\\
{\itshape \href{http://icecube.wisc.edu/collaboration/authors/icrc19_icecube}{http://icecube.wisc.edu/collaboration/authors/icrc19\_icecube}}\\
E-mail: \email{rcross2@ur.rochester.edu}, \email{afritz04@uni-mainz.de}, \email{sgriswol@ur.rochester.edu}
}

\abstract{
The IceCube Neutrino Observatory, which instruments 1$\,$km$^3$ of clear ice at the geographic South Pole, was mainly designed to detect particles with energies in the multi-GeV to PeV range.  Due to ice temperatures between $-20^\circ$C to $-43^\circ$C and the low radioactivity of the ice, the dark noise rates of the 5160 photomultiplier tubes forming the IceCube lattice are of order 500 Hz, which is particularly low for 10 inch photomultipliers. 
 Therefore, IceCube can extend its searches to bursts of $\mathcal{O}$(10\,MeV) neutrinos lasting several seconds, which are expected to be produced by Galactic core collapse supernovae. By observing a uniform rise in all photomultiplier rates, IceCube can provide a particularly high statistical precision for the neutrino rate from supernovae in the inner part of our Galaxy ($<$ 20 kpc). In this paper, the tools and the method to study potential obscured or failed core collapse supernovae in our Galaxy are presented. The analysis will be based on 3911 days of IceCube data taken between April 17, 2008 and December 31, 2018. 

\vspace{4mm}
{\bfseries Corresponding authors:}
Robert Cross$^{1}$, Alexander Fritz$^{2}$, \speaker{Spencer Griswold}$^{1}$\\
{$^{1}$ \itshape University of Rochester}\\
{$^{2}$ \itshape University of Mainz}

}

\FullConference{36th International Cosmic Ray Conference -ICRC2019-\\
		July 24th - August 1st, 2019\\
		Madison, WI, U.S.A.}

\begin{document}

\section{Introduction and Detection Principle}\label{sec:intro}

This contribution outlines the method for a 11-year search for core collapse supernovae within the Milky Way
that have been hidden optically or which failed to explode, as well as recent improvements in the simulation, detection and analysis of supernova candidates in IceCube.

\begin{figure}[ht]
  \centering
    \includegraphics[angle=0,width=0.46\textwidth]{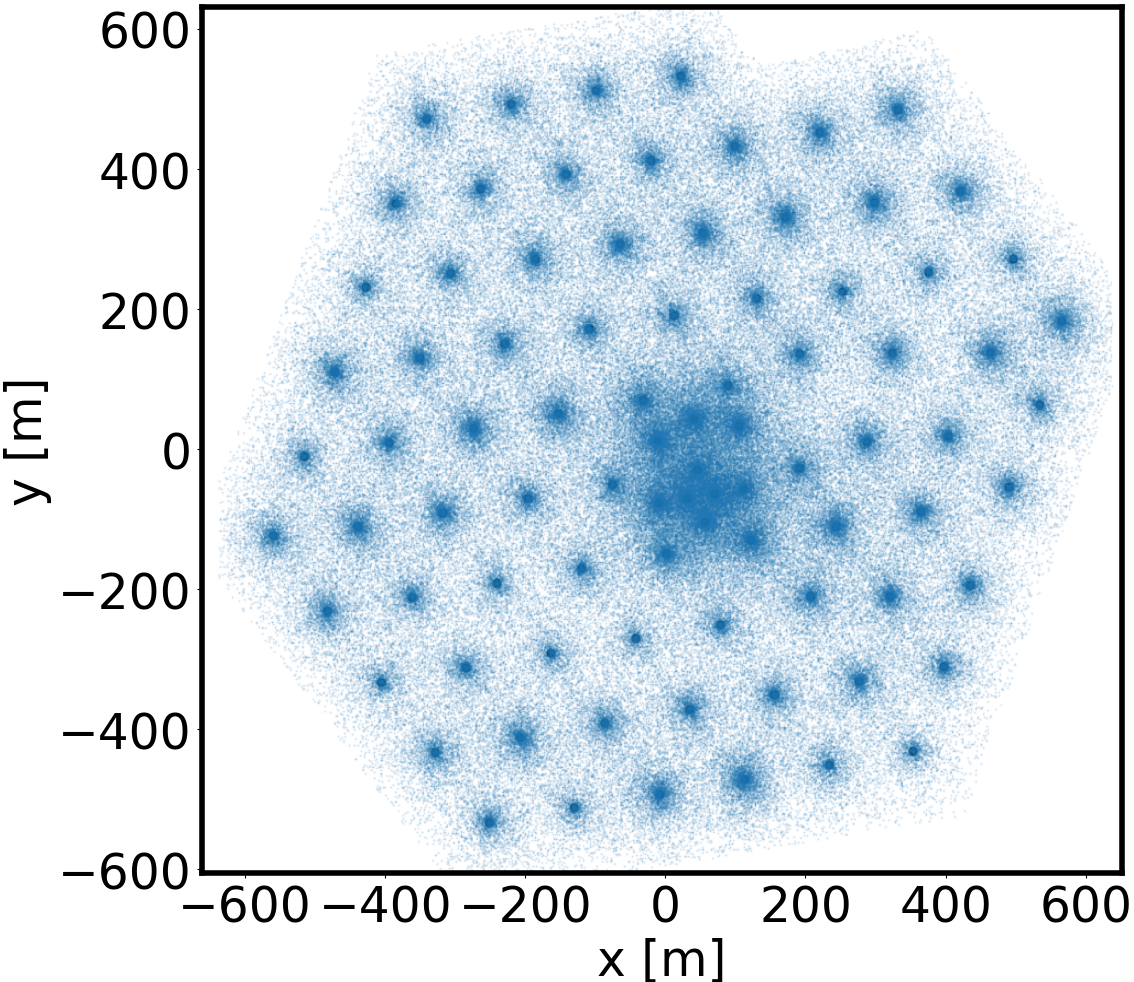}\hfill
  \includegraphics[angle=0,width=0.48\textwidth]{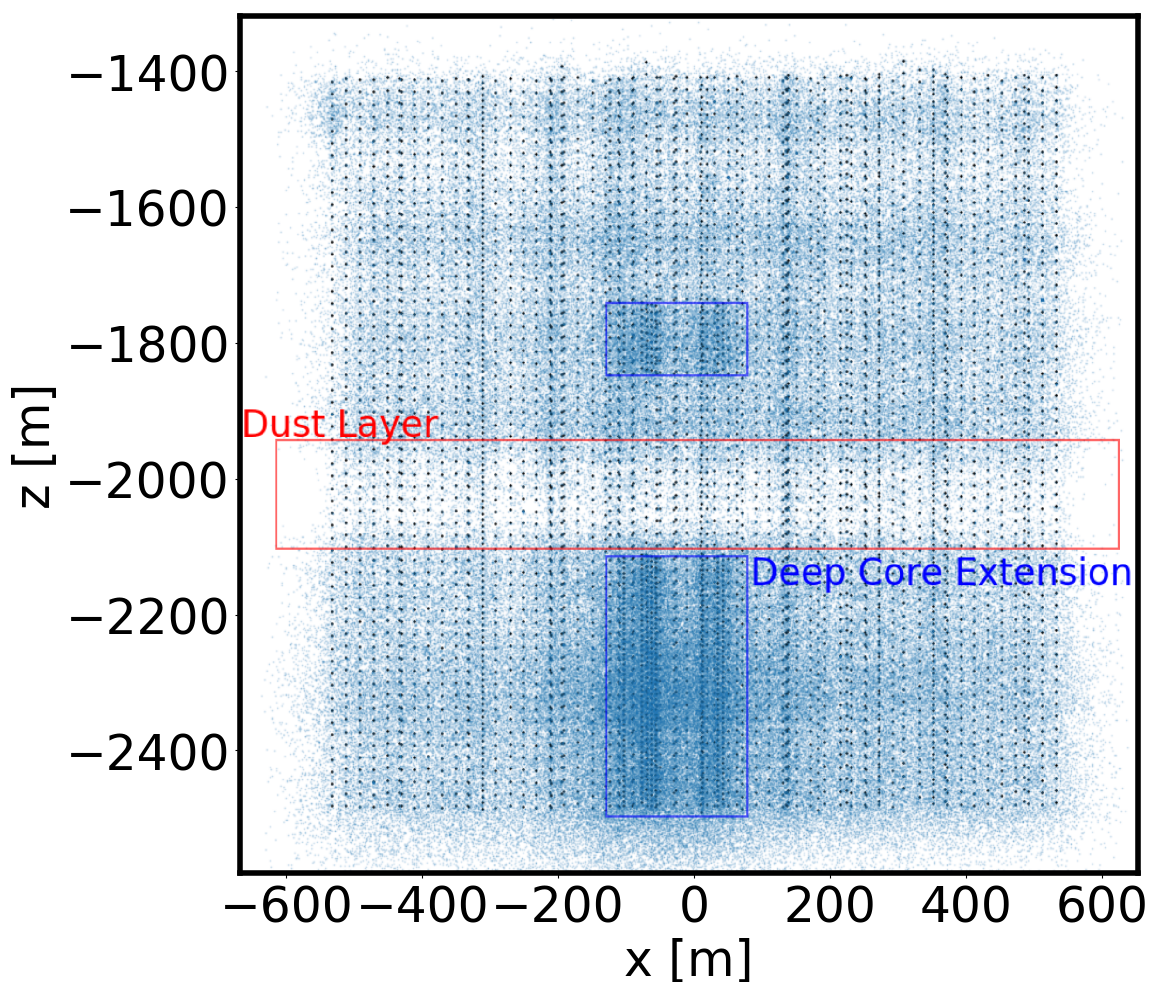}

  \caption{Top and side view of $\sim3.4\times10^5$ simulated supernova $\nu$ interaction vertices registered by IceCube DOMs. The dust layer between  -1950$\,$m and -2050$\,$m  and the denser DeepCore subarray are clearly visible.}\label{fig:GEANT}
\end{figure}

IceCube is a grid of 5160 digital optical modules (DOMs) embedded 1450 to 2450~m below the surface of the ice sheet at the geographic South Pole \cite{Achterberg:2006md}. The DOMs are deployed on 86 strings separated horizontally by 125~m, with the DOMs on each string spaced 17~m from their neighbors. IceCube is uniquely suited to monitor our Galaxy for supernovae due to its 1~km$^3$ volume and cold environment \cite{Halzen:1995ex}. In the inert $-43^\circ$C to $-20^\circ$C ice, the DOM noise rates average $\sim540$~Hz. The total cosmic ray muon event rate in the deep ice is $\sim5$~kHz. The optical absorption length in the ice exceeds 100~m, so each DOM effectively monitors several hundred m$^3$ of ice (see Fig.~\ref{fig:GEANT}).

A comprehensive survey of supernova physics is provided in \cite{Handbook:2017}, but in summary, the density inside a core-collapse supernova is sufficiently high to produce a thermal population of neutrinos of all species that diffuse out and eventually reach Earth. The neutrino energy spectrum is expected to peak between 10 and 20~MeV, depending on the mass and equation of state of the stellar progenitor. The inverse beta decay (IBD) process $\bar{\nu}_\mathrm{e}+\mathrm{p}\to \mathrm{n}+\mathrm{e}^+$, whose cross section depends approximately on the square of the anti-neutrino energy $E_{\bar{\nu}_\mathrm{e}}$, is the dominant supernova neutrino interaction in ice. Typically, only a single photon from each IBD reaches one of the DOMs.
During a supernova, the increase in photon counts in individual DOMs will not be statistically significant, but the collective increase in counts across all DOMs produces a strong signal~\cite{Abbasi:2011ss}. The signal does not allow for reconstruction of individual neutrino interactions or estimates of the energy, origin, and type of neutrino, but it provides detailed measurements of the neutrino flux versus time.

The reconstruction of supernova neutrino luminosity from the excess photon count rate $R(t)$ observed in IceCube proceeds as follows. Positrons with energy $E_{\mathrm{e}^+}$ radiate an average of $\expval{N_\gamma}\approx 178\cdot (E_{\mathrm{e}^+}/\mathrm{MeV})$ Cherenkov photons between 300~nm and 600~nm. 
The average number of photons depends on the cross section and positron track length and is thus roughly proportional to $E_{\bar{\nu}_\mathrm{e}}^3$. Assuming the shape of the neutrino spectrum can be characterized by a pinching parameter
$\alpha$ \cite{Keil:2003sw}, 
%
%
the average number of Cherenkov photons produced per positron will be $\expval{N_\gamma}\propto \expval{E_{\bar{\nu}_\mathrm{e}}}(\alpha+4)/(\alpha+1)$
with variance $\Var{(N_\gamma)}\propto \expval{N_\gamma}^2/(\alpha+4)$. If the progenitor is at distance $d$ from Earth and produces a neutrino number flux $\Phi_{\bar{\nu}_\mathrm{e}}$ (in m$^{-2}$~s$^{-1}$), the estimated energy luminosity is
\begin{align*}
%
%
L_{\overline{\nu}_\mathrm{e}}(t)\left[\frac{\mathrm{MeV}}{\mathrm{s}}\right] &= 4\pi d^2\expval{E_{\bar{\nu}_\mathrm{e}}(t)} \Phi_{\bar{\nu}_\mathrm{e}}(t) \\
&\approx  6.67\times 10^{52}\cdot R(t)
 \qty(\frac{d}{10\;\mathrm{kpc}})^2
 \qty(\frac{\expval{E_{\bar{\nu}_\mathrm{e}}(t)}}{15\;\mathrm{MeV}})^{-2} \frac{\left(1+\alpha(t)\right)^2}{\qty[2+\alpha(t)]\qty[3+\alpha(t)]}
 \;\;.
\end{align*}
Note that flavor mixing effects in the collapsing star can change the neutrino energy spectra and fluxes, modifying the expected rate in IceCube. 
 In the central $\sim 100$ km radius of the star, the neutrino density exceeds the electron density and $\nu$-$\nu$ coherent scattering sets in, leading to complex energy- and time-dependent flavor conversions. 
At larger radii, the neutrino flavor conversion is driven by coherent scattering on electrons with resonant enhancements at densities around $1000$ and $10$~g~cm$^{-3}$. Eventually, the neutrinos exit the star in their vacuum state $\nu_i$. Typically, the most conservative rate assumption is that no matter oscillations occur.

Construction of IceCube finished in 2011, and since 2015 the trigger-capable uptime of the detector has averaged 99.7\% around the clock. 
Due to the  non-Poissonian character of the dark noise in the IceCube DOMs \cite{Abbasi:2011ss}, the data acquisition system incorporates an artificial deadtime of $\tau=250~\mu\mathrm{s}$ to reduce the dark rate $R_\mathrm{dark}(t)$ by $\approx50\%$. The deadtime also lowers the detector count rate by a factor $0.87/(1+R_\mathrm{dark}(t)/N_\mathrm{DOM}\cdot \tau)$, where $N_\mathrm{DOM}$ is the number of participating optical modules.
%
%
%
DOM rates are counted in 1.6384~ms time bins. A dedicated online software system (SNDAQ) rebins the data to 2~ms and searches the data stream for collective rate increases characteristic of a supernova. SNDAQ computes a moving-average search for rate increases using fixed time bins of 0.5, 1.5, 4, and 10~s based on the typical timescales of features in the supernova neutrino light curve \cite{Abbasi:2011ss}. Since October 2018, the online search has been supplemented by a Bayesian Blocks algorithm, a dynamic self-learning histogramming method with variable bin widths \cite{Cross:2017hyw}. The Bayesian Blocks search provides a model-independent trigger for signals exceeding a duration of 0.5~s, with a trigger threshold that can be tuned to a chosen false positive rate.

Since the timing accuracy of the online monitor is limited to 2~ms, an improved readout system has operated since 2014 to buffer and extract the full DOM waveforms if triggered by a supernova candidate \cite{Baum:2013}. Since 2018, the automatic buffer has included triggers from the Supernova Early Warning System (SNEWS) \cite{Antonioli:2004zb} and LIGO-Virgo gravitational wave alerts \cite{deWasseige:2019icrc_gw}.

\section{Detector Simulation and Expected Performance}\label{sec:simulation}

Currently, three simulation schemes are used to estimate the expected rates in IceCube. In increasing order of speed and decreasing order of sophistication, they are: a GEANT-4 based simulation of individual supernova neutrino interactions in the ice and a GPU raytracer for Cherenkov photons; ASTERIA, a fast parameterized simulation of the detector response written in Python \cite{ASTERIA:2019}; and an implementation of the IceCube detector response in SNOwGLoBES \cite{Malmenbeck:2019icrc}, useful for quick comparisons of IceCube with other supernova detectors.

The GEANT simulation uses the IceCube offline simulation software and produces reconstructed events with an average rate of 15 events~s$^{-1}$; one of every 450 interactions in the sparsely instrumented IceCube and DeepCore volumes yields at least one registered Cherenkov photon. The corresponding positions of neutrino interactions are shown in
Fig.~\ref{fig:GEANT}.

\begin{figure}[t]
  \centering
  \includegraphics[width=\textwidth]{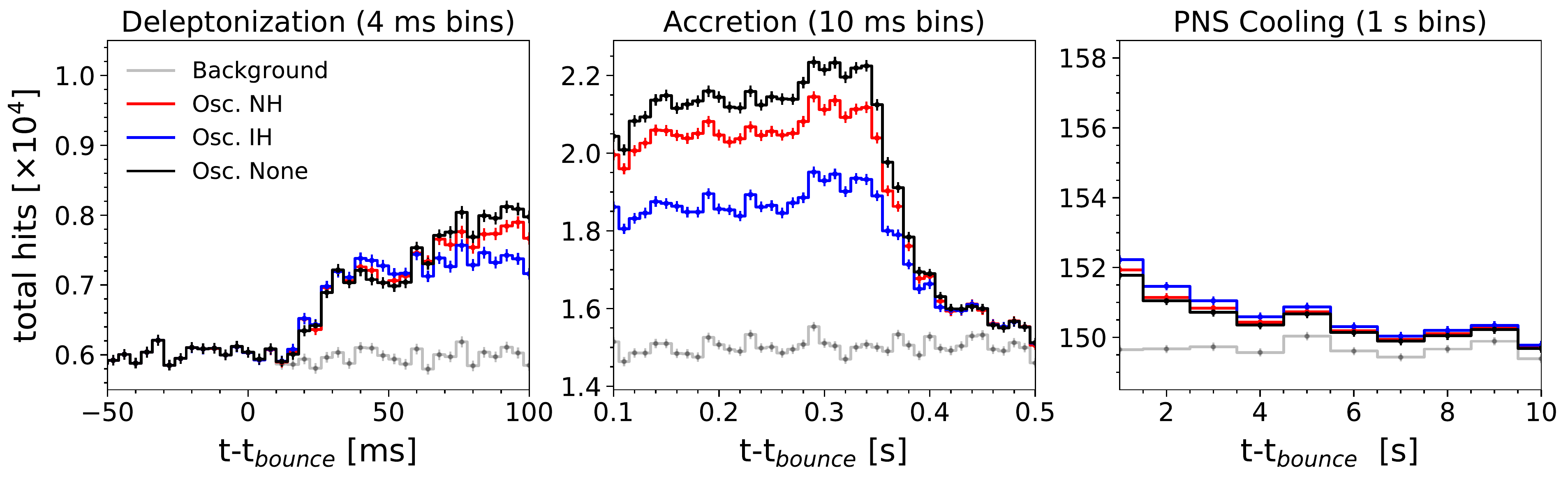}
  \caption{Simulated DOM hits in IceCube for a supernova from a $13$~M$_\odot$ star located $10$~kpc from Earth \cite{Nakazato:2012qf}, assuming several different neutrino oscillation scenarios \cite{ASTERIA:2019}.}
  \label{fig:hitsim}
\end{figure}

Using the effective volume of each DOM from the GEANT-4 simulation, ASTERIA is used to quickly model the detector response to different supernova neutrino emission models. Fig.~\ref{fig:hitsim} shows the expected hit rate in IceCube from a $13$~M$_{\odot}$, 0.02 metallicity progenitor \cite{Nakazato:2012qf} located 10~kpc from Earth. The signal in IceCube is dominated by the $\bar{\nu}_\mathrm{e}$ flux measured during the accretion phase of the explosion, but the detector is also sensitive to the cooling tail of the proto-neutron star.

The expected neutrino flux from a supernova contains significant systematic uncertainties which depend on the mass and equation of state of the progenitor and the simulation code used to generate the explosion. Predicted neutrino fluxes from a fixed distance can easily vary by a factor of 10 depending on the supernova simulation and progenitor model. The observed rates also depend on the assumed neutrino mass hierarchy, the influence of matter-induced oscillations in the Earth, the assumed progenitor distribution in the Milky Way as well as the codes used and symmetry assumptions made in the simulation. By comparison, detector systematic uncertainties are relatively small, dominated by the effective volume uncertainty of 12\%, the uncertainty in the positron track length of 5\%, and 3\% uncertainties each from the effect of artificial deadtime and knowledge of interaction cross sections~\cite{Abbasi:2011ss}.

\vspace{-10pt}
\begin{figure}[ht]
  \begin{minipage}[l]{0.525\textwidth}
    \includegraphics[width=0.95\textwidth]{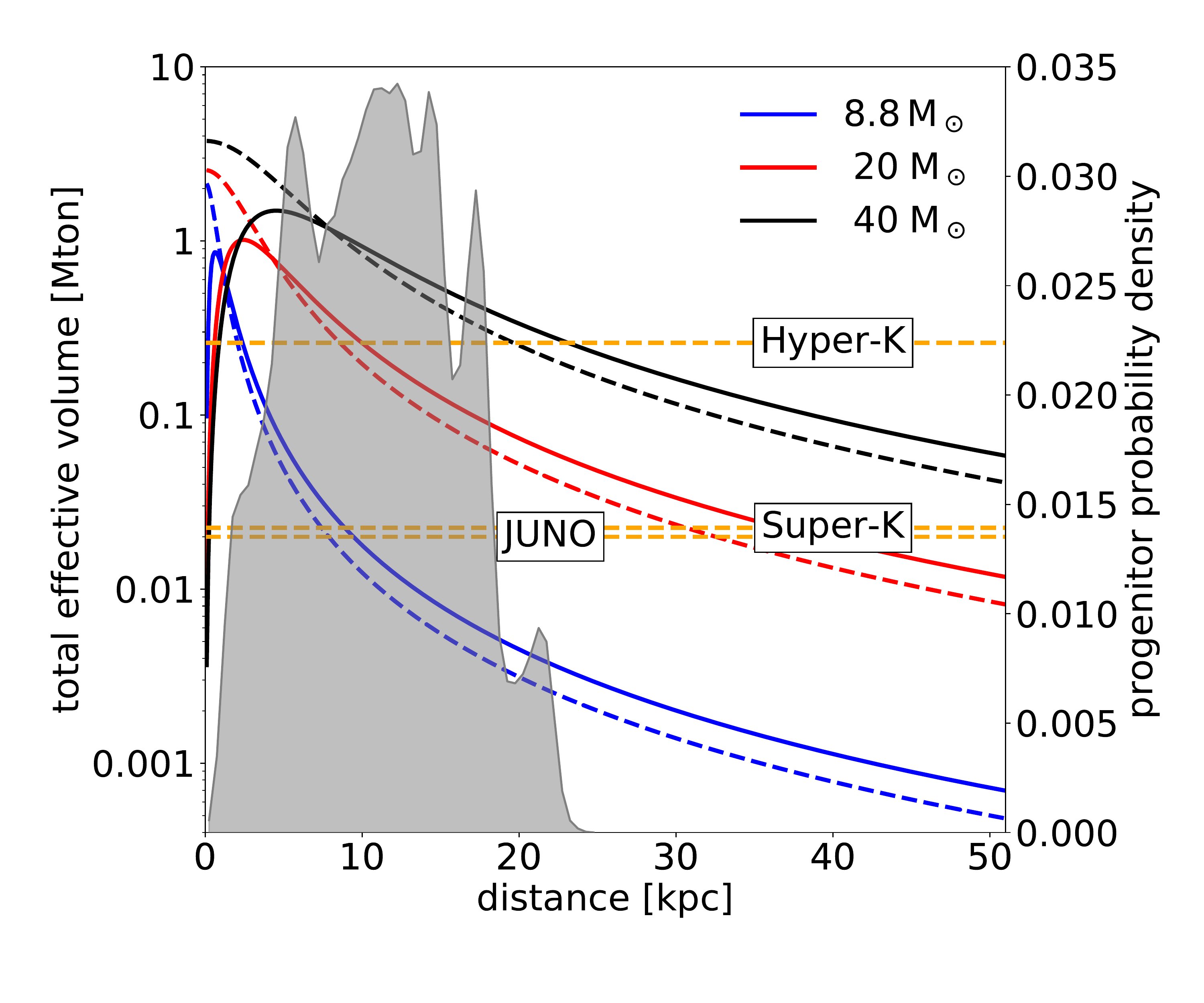}
  \end{minipage}
  \begin{minipage}[c]{0.475\textwidth}
    \vspace{-15pt}
    \caption{Comparison of effective total volumes of ideal and background free detectors for $8.8$~M$_\odot$ \cite{Huedepohl:2009wh}, $20$~M$_\odot$ \cite{Totani:1997vj}, and $40$~M$_\odot$ \cite{Sumiyoshi:2007pp} stellar models. The solid curves show the IceCube effective volume when the artificial deadtime of $250~\mu\mathrm{s}$ is applied; the dashed curves show the results without the deadtime. For comparison, the estimated fiducial volumes of Super-Kamiokande, JUNO and Hyper-Kamiokande are shown as dashed orange lines. An example progenitor probability density~\cite{Ahlers:2009ae} is shown in gray.}
    \label{fig:backgroundfree}
  \end{minipage}
\end{figure}
\vspace{-20pt}
It is instructive to compare the statistical resolution of time structures in the neutrino emission with that of an ideal background-free detector with 100\% detection efficiency for $\mathrm{e}^+$. Due to the IceCube dark rate, its precision depends on the relative number of signal hits registered in comparison to the background, i.e., on the distance, the luminosity, and the average neutrino energy in the time period studied. 
%
%
Fig.~\ref{fig:backgroundfree} shows a comparison of three models which deliver $\bar{\nu}_\mathrm{e}$ luminosities of  $2.9\times 10^{51}$, $3\times 10^{52}$, and $4.7\times 10^{52}\,$ erg within the first 1.5$\,$s from the explosions of 8.8~M$_\odot$, 20~M$_\odot$ and 40~M$_\odot$ progenitors. The average $\bar{\nu}_\mathrm{e}$ energies in this range are  12.4, 14.8, and 23.1$\,$MeV, respectively. Oscillations have not been taken into account. For comparison, orange lines corresponding to the fiducial volumes of Super-Kamiokande (22.5$\,$kt), JUNO (20$\,$kt)  and  the planned Hyper-Kamiokande experiment (260$\,$kt) are also drawn, as well as a progenitor probability density distribution~\cite{Ahlers:2009ae}. At larger distances, the IceCube artificial deadime improves the sensitivity (solid curves) while at low distances, the deadtime cuts into the signal, eventually reaching a rate of $1/\tau$. For progenitors that are not overly light and distances up to the center of our Galaxy, IceCube is the best instrument to detect neutrino flux variations during the accretion phase.

\begin{figure}[ht]
  \centering
  \includegraphics[width=0.47\textwidth]{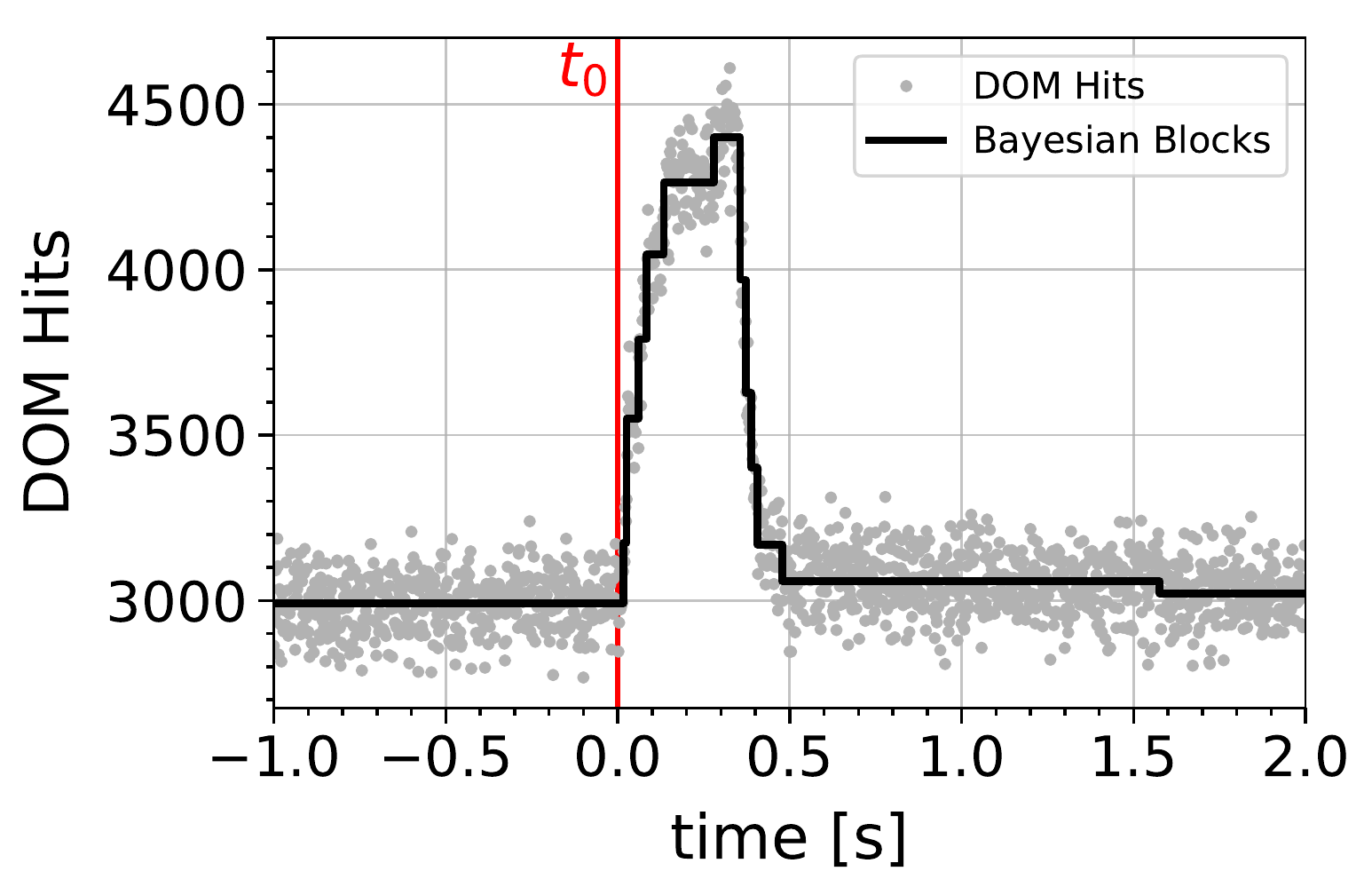}
  \hspace{5mm}
  \includegraphics[width=0.45\textwidth]{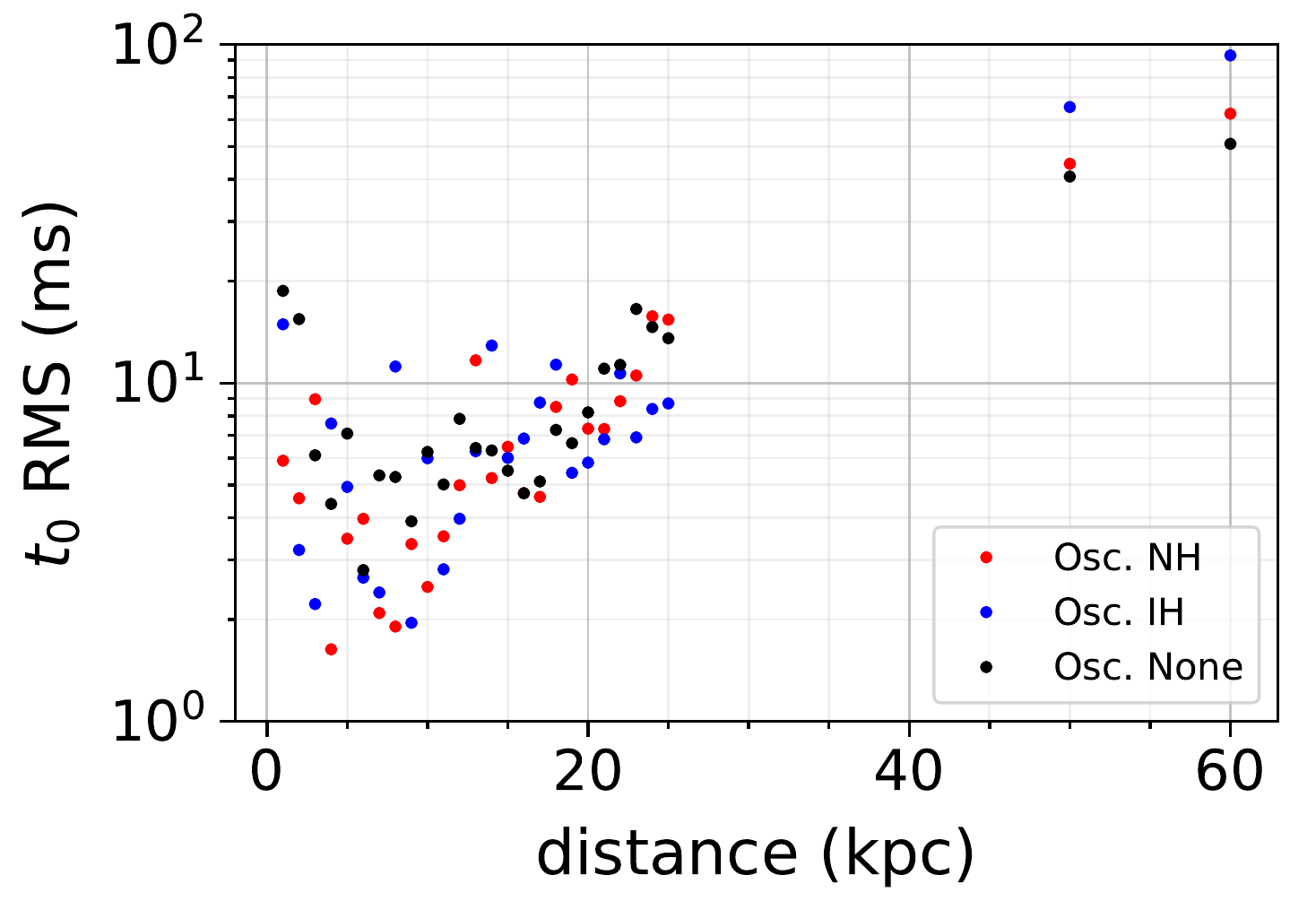}
  \caption{{\sl Left:} Changes in the rate of simulated DOM hits from a $13$~M$_\odot$ supernova \cite{Nakazato:2012qf} 10~kpc from Earth found by Bayesian Blocks \cite{Cross:2017hyw}. {\sl Right:} RMS error in the estimate of the start time $t_0$ vs. supernova distance.}
  \label{fig:bb}
\end{figure}

Figure~\ref{fig:bb} shows the simulated DOM hits in IceCube produced by a $13$~M$_\odot$ progenitor near the Galactic Center \cite{Nakazato:2012qf}. The Bayesian Blocks algorithm implemented in SNDAQ \cite{Cross:2017hyw} has identified statistically significant changes in the hit rate binned in 2 ms without any model assumptions about the underlying explosion. Of particular interest is the time $t_0$ when the signal is first detected above background, since this initial real-time determination can be combined with data from other neutrino telescopes to triangulate the location of the supernova \cite{Muhlbeier:2013gwa, Brdar:2018zds}. The RMS error in $t_0$ as a function of distance to the supernova, shown in the right panel of Fig.~\ref{fig:bb}, is $\sim3-4$~ms for a progenitor 10~kpc from Earth.

\section{Search for Galactic Supernovae}


The rate of Galactic stellar collapses, including optically obscured and failed supernovae which produce black holes, is estimated to be $1.7$ to $2.5$ per 100 years \cite{Giunti:2007ry}. 
The Baksan experiment (31.3 years of livetime) quotes a 90\% C.L. limit of < 7.4 core collapse supernovae per century within 20~kpc \cite{Petkov:2017cca}, and LVD (23.5 years of livetime) quotes a 90\% C.L. limit of < 9.8 per century within 25~kpc \cite{Vigorito:2017aqd}. Both LVD and Baksan determined their limits by adopting phenomenological models that were parameterized to fit the observation of SN1987A.


As discussed in Section~\ref{sec:simulation}, the neutrino emission from a core collapse supernova may vary by an order of magnitude depending on the mass and type of progenitor. It is therefore important to specify which models have been assumed for a supernova search with a neutrino detector. In this analysis, five core-collapse supernova simulations in spherical symmetry, covering a large spread of neutrino emission, were selected as benchmark models.

The minimal initial mass of a progenitor that can produce a supernova is $(8 \pm 1)$~M$_\odot$ \cite{Smartt:2009zr}. For such low masses, the collapse is induced by electron capture in a degenerate O-Ne-Mg core. %
%
%
The collapse of a 8.8~M$_\odot$ O-Ne-Mg progenitor (the ``H\"udepohl model'' \cite{Huedepohl:2009wh}) is an example where 1D simulations yield neutrino-powered supernova explosions. This low mass model, with a total emitted energy of $1.7\times10^{53}$~erg and $\expval{E_{\bar{\nu}_\mathrm{e}}}\approx11.6$~MeV, represents a  conservative lower limit on the expected $\bar{\nu}_\mathrm{e}$ luminosity and energy spectrum. A second model, corresponding to an $11.2$~M$_\odot$ progenitor \cite{Mirizzi:2015eza}, yields a total emitted energy of $2.1\times10^{53}$~erg and $\expval{E_{\bar{\nu}_\mathrm{e}}}\approx12.9$~MeV. Our third and fourth benchmark models use higher mass progenitors: a $27$~M$_\odot$ star that yields $3.3\times10^{53}$~erg and $\expval{E_{\bar{\nu}_\mathrm{e}}}\approx13.7$~MeV \cite{Mirizzi:2015eza}; and a 1D model with a forced explosion of a $30$~M$_\odot$ progenitor that yields $1.97\times10^{53}$~M$_\odot$ and $\expval{E_{\bar{\nu}_\mathrm{e}}}\approx16.2$~MeV \cite{Nakazato:2012qf}. For stellar masses $>25$~M$_\odot$, gravitational collapse may lead to a limited explosion followed after $\sim1$~s by the formation of a black hole; stars $>40$~M$_\odot$ are not expected to explode at all. Such a ``Black Hole model'' has two distinguishing features: an average energy $\expval{E_{\bar{\nu}_\mathrm{e}}}$ roughly twice as large as in exploding stars, due to the continual accretion of material on the core; and a sharp cutoff in the neutrino flux after $\sim1$~s \cite{Sumiyoshi:2007pp}. For this analysis, we assume a 30$\,\mathrm{M_\odot}$ progenitor and a hard equation of state. 


More than 80\% of supernovae may be obscured by dust and would thus not be optically visible~\cite{Mattila:2012zr}. The search method should therefore not depend on external information. In IceCube, SNDAQ computes a moving average test statistic $\xi=\Delta\mu/\sigma_{\Delta\mu}$, where $\Delta\mu$ is the most likely collective rate deviation of all DOM hit rates from their running average. The uncertainty $\sigma_{\Delta\mu}$ is calculated from the data themselves, thus accounting for non-Poissonian behaviour in the dark rates. The test statistic $\xi$ (also termed pre-trial significance) should be distributed as a zero-mean unit Gaussian if no correlations are present in the rates. The calculation was done in overlapping 1.5~s time intervals using 500~ms time steps. The largest $\xi$ value in a 10~s time interval was selected.     

\begin{figure}[ht]
  \begin{minipage}[l]{0.65\textwidth}
    \includegraphics[width=0.95\textwidth]{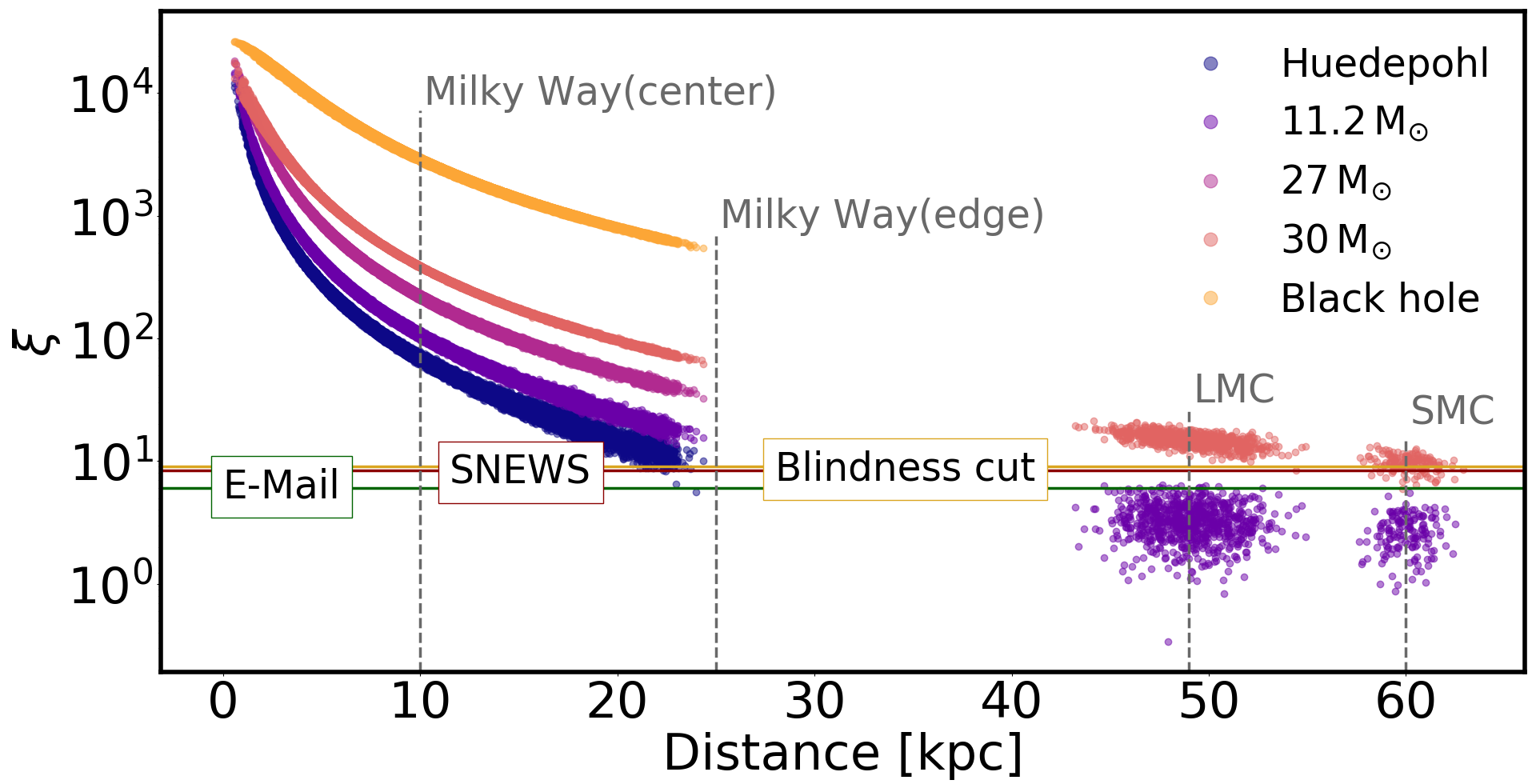}
  \end{minipage}
  \begin{minipage}[c]{0.35\textwidth}
    \vspace{-10pt}
    \caption{Test statistic $\xi$ (in units of Gaussian $\sigma$) vs. progenitor distance, simulated with ASTERIA and SNDAQ for the five models discussed in this paper: the O-Ne-Mg core from H\"udepohl et al. \cite{Huedepohl:2009wh}; an $11.2$~M$_\odot$ star \cite{Mirizzi:2015eza}; a $27$~M$_\odot$ star \cite{Mirizzi:2015eza}; a forced explosion of a $30$~M$_\odot$ star \cite{Nakazato:2012qf}; and a failed supernova which formed a black hole \cite{Nakazato:2012qf}.}
    \label{fig:reach}
  \end{minipage}
\end{figure}


Starting with a data set encompassing 3911 days from April 17, 2008 to December 31, 2018, several requirements are introduced to select high quality data. Short runs ($<10$~min), runs taken with calibration light sources, and runs with an incomplete detector configuration were discarded, removing 2.6\% of the data. Livetimes ranging between 95.8\% and 98.3\% were achieved between 2013 and 2019, while the livetime from 2008 to 2012 are $\sim90\%$. The total livetime after quality cuts corresponds to 3670 days.  

The parameterized simulation \cite{ASTERIA:2019}, verified to produce the same results as the GEANT-4 Monte Carlo, was used to simulate 18000 supernovae using the 8.8$\,\mathrm{M_\odot}$ ~\cite{Huedepohl:2009wh}, 11.2$\,\mathrm{M_\odot}$ ~\cite{Mirizzi:2015eza}, 27$\,\mathrm{M_\odot}$~\cite{Mirizzi:2015eza}, 30$\,\mathrm{M_\odot}$ ~\cite{Nakazato:2012qf} and black hole~\cite{Sumiyoshi:2007pp} progenitor models. The progenitors' distances were distributed according to the parameterizations of the radial distribution of Galactic structure published by Mirizzi et al.~\cite{Mirizzi:2006xx} and Ahlers et al.~\cite{Ahlers:2009ae}. The simulated hit rates were added to uniformly sampled DOM rates recorded by SNDAQ between 2008 and 2018.

\begin{figure}
  \centering
  \includegraphics[angle=0,width=0.50\textwidth]{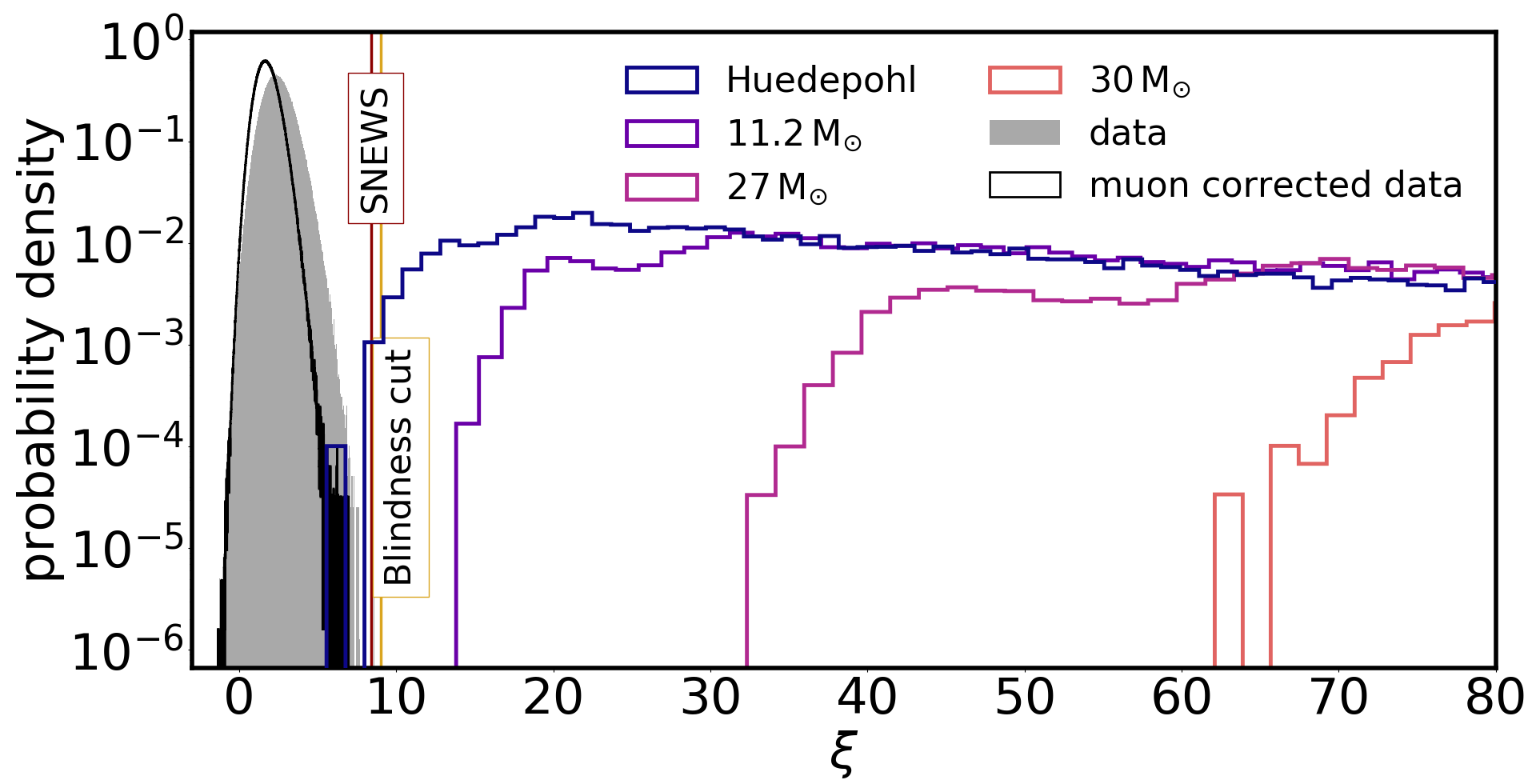}
  \includegraphics[angle=0,width=0.49\textwidth]{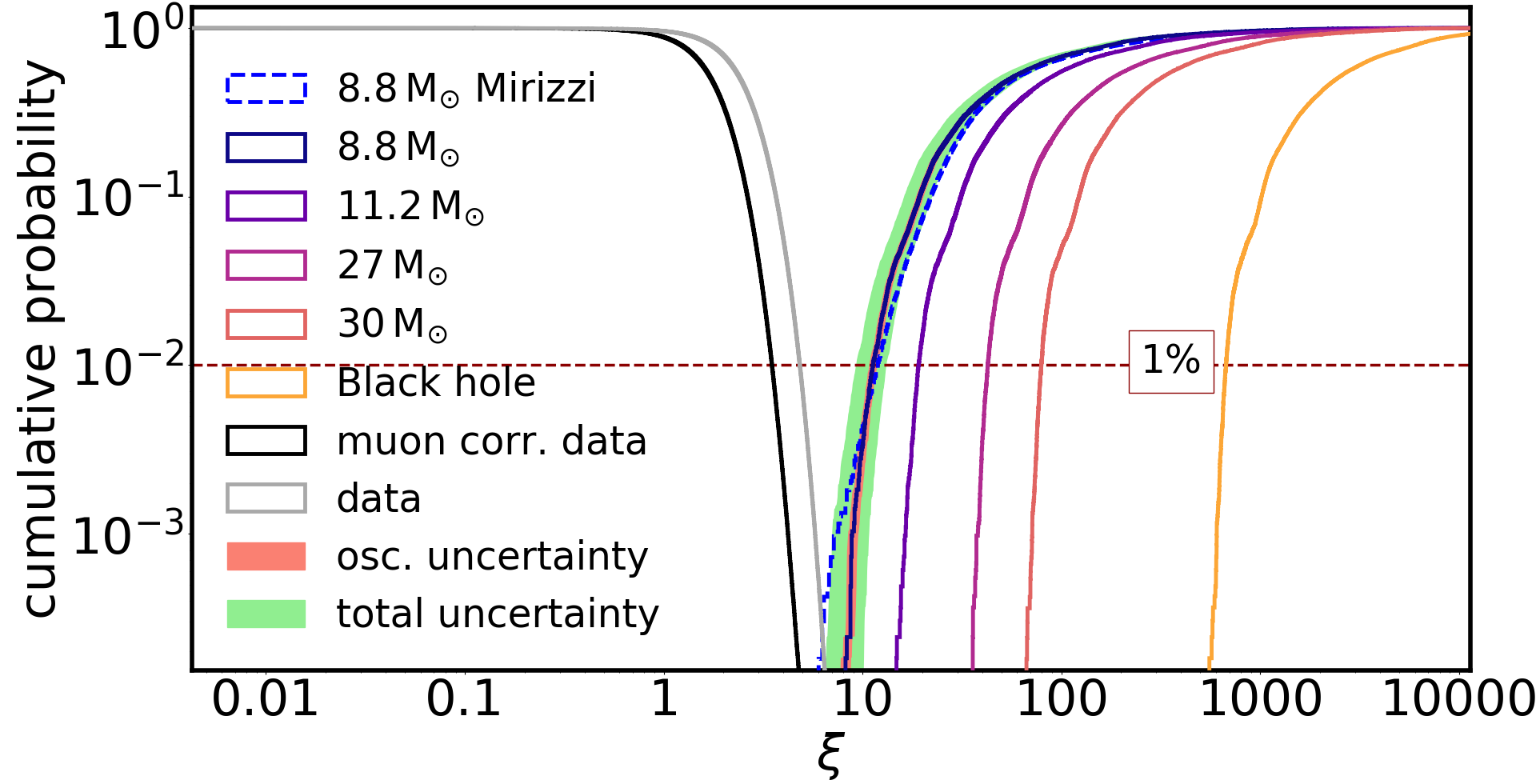}
  \caption{{\sl Left:} 
The distribution of the test statistic $\xi$ is shown for data taken in 2014 with and without muon subtraction as well as for four supernova models. A slight overlap occurs only for the model with the lowest progenitor mass~\cite{Huedepohl:2009wh}. {\sl Right:} Fraction of supernovae missed in our Galaxy by imposing a cut on the test statistic $\xi$. The colored bands show the effects of oscillations and the 14\% systematic detector uncertainty for the case of the H\"udepohl model. Also shown is the effect of the chosen progenitor radial distribution model   (\cite{Mirizzi:2006xx} (solid line), \cite{Ahlers:2009ae} dashed line).
The horizontal line indicates the cut value, where 99\% of all supernovae are retained. 
Vertical lines indicate the blindness as well as the cuts that are used for dissemination by SNEWS. }
  \label{fig:significance}
\end{figure}

Fig.~\ref{fig:reach} displays the simulated $\xi$ distribution for the detection of supernovae as function of distance. The plot also includes the positions of the Magellanic Clouds at assumed distances of 49$\,$kpc and 60$\,$kpc, respectively. 
Fig.~\ref{fig:significance} shows the distributions of the test statistic $\xi$ for data taken in 2014 with and without muon subtraction and for the five core collapse models under consideration. The left and right panels shows the probability and cumulative distributions, respectively. Note that in the left panel, the entire distribution of $\xi$ for the black hole model lies beyond the right edge of the plot, due to the high luminosity and average energy produced by this model during the accretion phase of the collapse. The effects of systematic uncertainties are shown for the H\"udepohl model in the right plot. To err on the side of caution, we will assume the neutrino mass hierarchy to be normal and apply the progenitor distribution of~\cite{Ahlers:2009ae}.  
The distribution narrows substantially when hits from atmospheric muons are subtracted (see dark-blue histograms in Fig.~\ref{fig:significance}). This occurs because cosmic-ray air showers produce muon bundles that trigger many optical modules. The muon rate depends on the season. 
While the average contribution to the count rates of individual optical modules is  only 3\%, these hits are correlated across the detector, broadening the distribution of $\xi$. Note that muon-corrected distributions were not used here; we will turn to them in future work 
to reduce false alerts in the study of light  progenitors and supernovae in the Magellanic Clouds.


So far, only data with $\xi$ values below $9\,\sigma$ ($7\,\sigma$) for data uncorrected (corrected) for cosmic ray muons have been studied. Once the data are unblinded -- and in case we observe no signal -- an upper limit will be provided that is valid for 99\,\% of all Galactic core collapse supernovae with neutrino fluxes equal or higher than in the conservative 8.8$\,\mathrm{M_\odot}$  H\"udepohl model. With the systematic uncertainties included so far, this requirement would correspond to a cut at $\xi > 9.3\,\sigma$. 
\section{Conclusion and Outlook}
We set up a search for neutrinos from core collapse supernovae in our Galaxy using IceCube data taken between April 2008 to December 2018 and discussed the required analysis and simulation tools. In the future, the analysis will be further improved by extending the search to the Magellanic Clouds and employing improved noise reduction techniques. As part of the approved IceCube Upgrade, multi-PMT modules~\cite{Classen:2019icrc} will be deployed and low-noise wavelength-shifting sensors~\cite{Peiffer:2017icrc} will be tested which have the potential to increase the distance reach and improve the spectral sensitivity. 

%
%



\bibliographystyle{ICRC}
\bibliography{references}

\clearpage

\end{document}